\documentclass[numbers]{assets/matterlab}
\usepackage{amsmath,amssymb,mathtools}
\usepackage{xurl}
\usepackage{float}
\usepackage{algorithmic}
\floatstyle{ruled}
\newfloat{algorithm}{tbp}{loa}
\floatname{algorithm}{Algorithm}
\floatstyle{plain}
\usepackage[noabbrev,nameinlink]{cleveref}
\hypersetup{pdftitle={MAPPA: Symmetry-Aware Atom Mapping through Continuous Fragment Growth},pdfauthor={Yunheng Zou and coauthors}}
\title{MAPPA: Mapping of Atoms by Progressive Pattern Assembly with Molecular Symmetry}
\author[2,6,8]{Yunheng Zou}
\author[10]{Olalla Nieto Faza}
\author[11]{Shifa Hussain}
\author[2,5,\dagger]{Varinia Bernales}
\author[1,2,3,4,5,6,7,8,9,\dagger]{Al\'an Aspuru-Guzik}
\affiliation[1]{Department of Chemistry, University of Toronto, 80 St. George St., Toronto, ON M5S 3H6, Canada}
\affiliation[2]{Department of Computer Science, University of Toronto, 40 St George St., Toronto, ON M5S 2E4, Canada}
\affiliation[3]{Department of Materials Science \& Engineering, University of Toronto, 184 College St., Toronto, ON M5S 3E4, Canada}
\affiliation[4]{Department of Chemical Engineering \& Applied Chemistry, University of Toronto, 200 College St., Toronto, ON M5S 3E5, Canada}
\affiliation[5]{Acceleration Consortium, 700 University Ave., Toronto, ON M7A 2S4, Canada}
\affiliation[6]{Vector Institute for Artificial Intelligence, W1140-108 College St., Schwartz Reisman Innovation Campus, Toronto, ON M5G 0C6, Canada}
\affiliation[7]{Canadian Institute for Advanced Research (CIFAR), 661 University Ave., Toronto, ON M5G 1M1, Canada}
\affiliation[8]{NVIDIA, 431 King St. W \#6th, Toronto, ON M5V 1K4, Canada}
\affiliation[9]{Institute of Medical Science, 1 King's College Circle, Medical Sciences Building, Room 2374, Toronto, ON M5S 1A8, Canada}
\affiliation[10]{Department of Organic Chemistry, Universidade de Vigo, Spain}
\affiliation[11]{Department of Biology, University of Toronto Mississauga, 1867 Inner Circle Rd., Mississauga, ON L5L 1C6, Canada}
\contribution[\dagger]{Principal investigators}
\abstract{Atom-to-atom mapping traces atom identities between molecular endpoints, supporting reaction analysis and pathway construction. Yet the same endpoints can admit multiple correspondences, and selecting a single answer can conceal alternatives with different chemical implications. Here we introduce Mapping of Atoms by Progressive Pattern Assembly (MAPPA), a framework that discovers and retains alternative atom mappings. It uses shared molecular fragments and molecular symmetry to represent these alternatives compactly, allowing their bond changes to be examined and supporting subsequent chemistry-aware selection. On the Golden benchmark (Golden), MAPPA had the highest reference coverage among the methods that we evaluated. With its standard settings, it recovered 99.03\% of the Golden and 99.63\% of the EnzymeMap reference mappings, with matching and decoding together averaging 3.19 and 13.16 CPU seconds per reaction, respectively. A broader search raised Golden coverage to 99.51\%. These results make broad exploration of mapping alternatives practical, providing a foundation for studying reaction ambiguity and incorporating chemical knowledge into mapping selection.
}
\date{November 8, 2026}
\metadata[Status]{Preprint}
\begin{document}
\maketitle
\clearpage
\section{Introduction}

A chemical process, from a flame to the cycle of a catalyst, is a network of elementary steps~\citep{unsleber2020exploration}. Once every step is known, with its intermediates and its transition state, the kinetics of the whole process can be modelled. This holds for catalysis as it does for atmospheric chemistry and combustion~\citep{steiner2022autonomous,meuwly2021machine}. For most of the history of chemistry these networks were drawn by hand, one mechanism at a time. Over the last decade, programs have begun to explore them automatically and at a much larger scale~\citep{simm2019exploration,baiardi2022expansive}. The costly part of each step is its transition state: thermodynamics needs only the energy minima, whereas kinetics also needs the saddle point between them~\citep{rappoport2014complex}. A large family of searches, the two-sided methods such as the nudged elastic band~\citep{henkelman2000}, start from both ends of the step, the reactant and the product. These methods carry a requirement that is easy to overlook: the program must be told which atom of the reactant becomes which atom of the product~\citep{de2019,vaucher2018curve}. De and co-workers stated the difficulty plainly. One could try every assignment, but their number grows as the factorial of the number of atoms of each element~\citep{de2019}: ten carbon atoms alone can be matched in more than three million ways. Several machine-learned generators of transition states take the same matched pair as their input~\citep{kim2024tsdiff,zeng2026transts}, and one was designed to do without it~\citep{duan2023oareactdiff}. This assignment is called atom-to-atom mapping (AAM). The general problem is NP-hard~\citep{jaworski2019mapping}, and it is not solved in practice: on a curated benchmark, the best of the tools compared mapped 84\% of the reactions as chemists did~\citep{lin2022}, and earlier methods mapped more than 5\% of a widely used set of patent reactions incorrectly~\citep{chen2024}. The quality of a reaction network therefore depends on the quality of its mappings.

A chemist who is unsure which atom went where does not guess. She draws two or three mechanisms and lets further evidence and calculations decide which one was right. We wanted a mapper that works in the same way. MAPPA returns several mappings for each reaction, and in our tests the one that a chemist would choose is almost always among them. The choice of the right atom-to-atom mapping is left for later, when more of the chemistry is known.

\begin{figure}[!b]
\centering\includegraphics{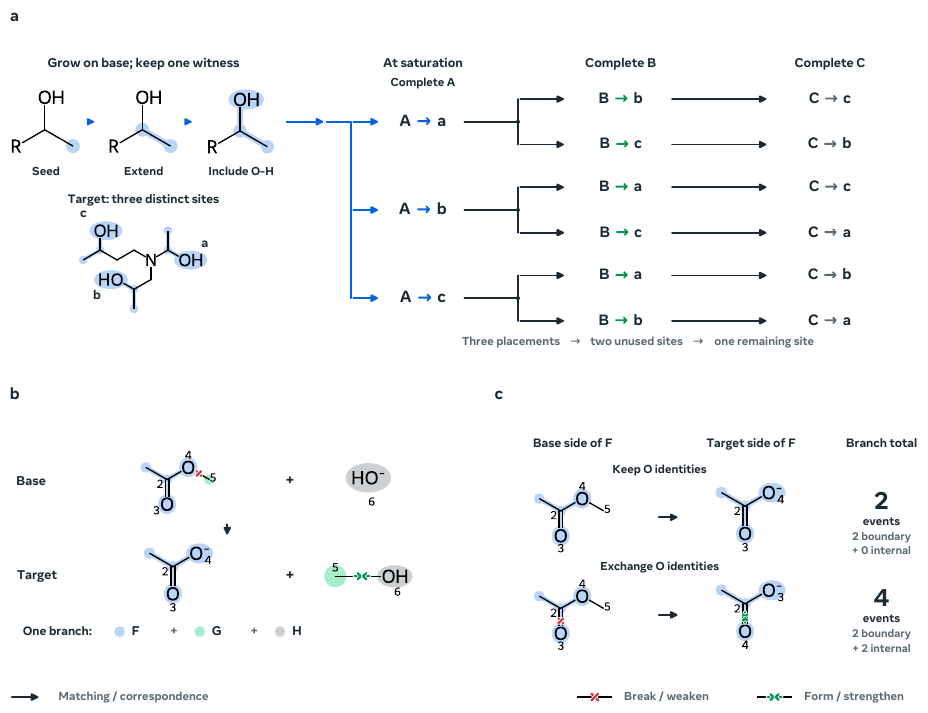}
\caption{\textbf{Fragment-based exploration of atom mappings with MAPPA.} (a) Greedy growth and conditional placements. A, B and C are successive local alcohol fragments, and R lies outside the matched motif. The C--C and C--O bonds and every attached H are checked. (b) Colors pair the base and target atom sets of fragments F, G and H. The branch has two boundary events: bond 4--5 breaks and bond 5--6 forms. (c) An automorphism of F's binary connectivity permits O3 and O4 to exchange. Original bond orders then give two internal events, added to the two boundary events in b. Both endpoints are shown; unhighlighted C5 is outside F. Numbers follow base atom identities. Red marks indicate cleavage or weakening, green marks formation or strengthening. Constructed examples are detailed in \Cref{sec:algorithm-details}.}
\label{fig:algorithm}
\end{figure}

A mapping assigns a correspondence between the atoms of the two molecular endpoints of a reaction, its reactants and its products. It describes the net evolution of atom identities through a chemical transformation, leaving the intervening elementary steps unspecified. Once this correspondence is established, changes in bonding can be identified explicitly (\Cref{fig:algorithm}). We call each bond that forms, breaks or changes order a \emph{bond event}. These assignments underpin reaction-center identification and the extraction of transformation templates used in reaction prediction and retrosynthesis~\citep{schwaller2021,chen2024}. Reliable mapping therefore matters both for interpreting individual reactions and for building the datasets and models used to study chemical reactivity.

\begin{figure}[!htbp]
\centering\includegraphics{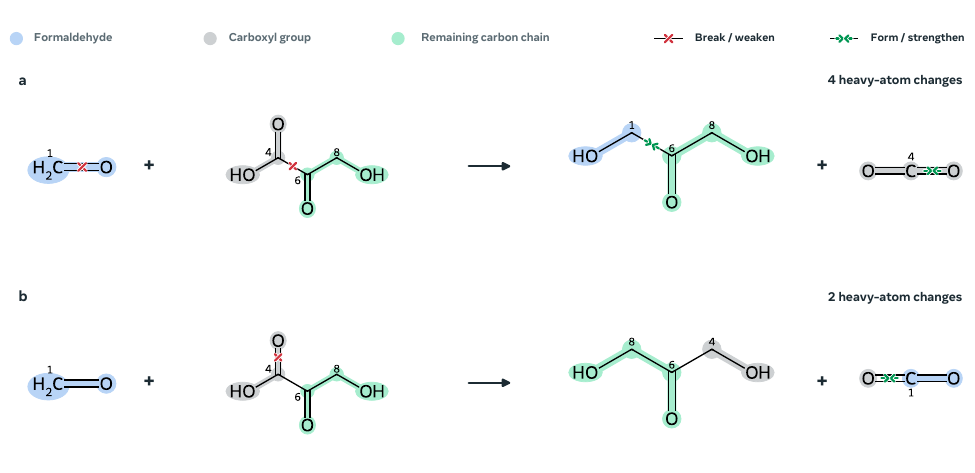}
\caption{\textbf{Distinct atom origins can share the same total event count.} Two saved MAPPA heavy-atom correspondences for EnzymeMap case 13175 retain all nine heavy atoms and the complete composition C$_4$H$_6$O$_5$. (a) The annotated mapping: formaldehyde carbon 1 enters the three-carbon molecule, while carboxyl carbon 4 becomes CO$_2$. One bond breaks, one forms, one weakens and one strengthens. (b) The alternative mapping exchanges these carbon origins. One bond breaks and one forms. The drawings show four and two heavy-atom bond changes, respectively; colors track atom origins and numbers identify the carbons. Minimizing hydrogen assignments separately for these fixed heavy-atom mappings adds two and four events, respectively, giving six total events in either case. The two correspondences are therefore tied under event counting. Hydrogen identities are not annotated by EnzymeMap. These correspondences do not establish reaction pathways. Provenance and counting details are given in \Cref{sec:alternatives-provenance}.}
\label{fig:alternatives}
\end{figure}

Endpoint structures, however, do not generally determine a unique correspondence. Molecular symmetry can make several atom labels structurally equivalent, while other assignments imply different atom origins and bond changes. Favoring fewer bond events narrows the choices, but it need not select the mapping annotated in a reference dataset, which we call the \emph{reference}. In an example from the EnzymeMap collection of enzymatic reactions~\citep{heid2023enzymemap} (\Cref{fig:alternatives}), the reference places the formaldehyde carbon in the three-carbon molecule, whereas another correspondence assigns it to CO$_2$. The two correspondences differ in their heavy-atom bond events, four and two, but both require six bond events in total when the hydrogen assignments are optimized separately for each. Both also account for the complete endpoint composition. Event counts therefore leave distinct atom origins unresolved in this example. Retaining alternatives allows subsequent interpretation to draw on additional chemical knowledge.

AAM methods can be organized by the output they seek to provide (\Cref{tab:aam-objectives}). \emph{Alternative-generating methods} retain several correspondences for subsequent interpretation or selection. SLAPMapper (SLAP) provides such alternatives through approximate graph matching, combining iterative refinement of atom labels with sequential linear assignment~\citep{koda2025}. \emph{Prediction-oriented methods} aim to select a preferred correspondence. RXNMapper derives assignments from attention patterns in a reaction language model trained on unannotated reactions~\citep{schwaller2021}; LocalMapper learns from chemist-labeled examples through a human-in-the-loop procedure~\citep{chen2024}. These methods bring information from reaction data into the selection of atom correspondences. \Cref{sec:previous} reviews these methods and the earlier ones.

\begin{table}[!htbp]
\centering\small
\caption{\textbf{A taxonomy of AAM methods by output objective.} Representative methods illustrate the distinction between retaining alternatives and selecting a preferred answer.}
\label{tab:aam-objectives}
\begin{tabular}{@{}>{\raggedright\arraybackslash}p{.18\linewidth}>{\raggedright\arraybackslash}p{.38\linewidth}>{\raggedright\arraybackslash}p{.38\linewidth}@{}}
\toprule
 & \textbf{Alternative-generating} & \textbf{Prediction-oriented}\\
\midrule
Objective & Preserve candidate correspondences for later interpretation or selection. & Identify the most plausible correspondence.\\
Output & Several mappings, stored individually or as compact families. & One preferred mapping.\\
Examples & Enumeration of optimal mappings~\citep{korner2008its,first2012ilp,ali2025}; SLAPMapper~\citep{koda2025}; MAPPA (this work). & RXNMapper~\citep{schwaller2021}; LocalMapper~\citep{chen2024}.\\
Evaluation & Reference coverage: is the annotated mapping retained? & Top-1 accuracy: does the selected mapping match the annotation?\\
\bottomrule
\end{tabular}
\end{table}

The two objectives can be combined by placing a selector after an alternative-generating method. A selector might favor fewer bond events, apply chemical rules, or use a learned ranking. This arrangement separates the ability to discover a useful correspondence from the ability to choose it. \emph{Reference coverage}, the share of reactions whose reference is among the returned mappings, measures the former. Top-1 accuracy, the share of reactions whose single selected mapping is the reference, measures the latter. A single predicted mapping can also be evaluated as a candidate set of size one, providing a common coverage criterion across the categories.

\begin{samepage}
For alternative generation, the computational challenge is to retain diverse correspondences without repeatedly searching and analyzing their shared structure. Choices among repeated groups can combine into many complete assignments, even when large parts of those assignments are identical. Enumerating each mapping separately repeats work and obscures which decisions distinguish the alternatives. An efficient representation must preserve those decisions while making their consequences for bond changes accessible. This need motivates organizing the search around shared molecular fragments and their allowed rearrangements.
\par\end{samepage}

Here we introduce Mapping of Atoms by Progressive Pattern Assembly (MAPPA), an alternative-generating mapper that treats atom correspondence as a constraint-search problem. \emph{Mappa} is also the Italian word for map. Greedy growth discovers fragments compatible with both endpoints; placing one fragment restricts the choices available to the remaining atoms. Molecular symmetry represents rearrangements within these constraints compactly. A matching stage discovers fragment combinations, and a decoding stage relates their alternatives to bond events, reusing calculations across shared fragments. On the Golden benchmark (Golden) of 1,851 reactions~\citep{lin2022} and on EnzymeMap~\citep{heid2023enzymemap}, runs with the standard settings reached a reference coverage of 99.03\% and 99.63\%, with matching and decoding together averaging 3.19 and 13.16 CPU seconds per reaction, respectively. A broader search raised Golden coverage to 99.51\%. We also examined how many alternatives remain under a minimum-event preference, connecting coverage and computational cost to the subsequent task of choosing a mapping.

\section{Previous work}
\label{sec:previous}

Automatic atom mapping is almost fifty years old. In this section we follow the principles by which earlier methods selected a mapping, and we say what MAPPA keeps from each.

\paragraph{Fewest bond changes.} The oldest principle is that a reaction changes as few bonds as it can. Jochum, Gasteiger and Ugi stated it in 1980 as the principle of minimum chemical distance~\citep{jochum1980pmcd}. The first programs that detected the reaction site, and those that searched for the largest substructure common to reactants and products, date from the same period~\citep{lynch1978detection,mcgregor1981mcs,funatsu1988reactionsite,moock1988reaccs}. Later work turned the principle into an optimization problem. Akutsu showed that the problem is NP-hard in general and that it can be solved in polynomial time for wide classes of enzymatic reactions~\citep{akutsu2004extraction}. Heinonen and co-workers found the exact minimum with an A* search~\citep{heinonen2011astar}. First, Gounaris and Floudas, and Latendresse and co-workers, found it by integer linear programming~\citep{first2012ilp,latendresse2012mwed}. The Reaction Decoder Tool, one of the mappers that we evaluate, also returns the mapping with the fewest modified bonds~\citep{rahman2016,preciatgonzalez2017comparative}. The principle has known exceptions. Jaworski and co-workers built their mapper on hand-written reaction rules because pericyclic reactions, 1,2-rearrangements and metathesis do not change the fewest bonds~\citep{jaworski2019mapping}. MAPPA therefore keeps the principle as a preference: it retains the mappings near the minimum, and not the minimum alone.

\paragraph{Learning from reaction data.} A second line of work learns the mapping from examples. RXNMapper reads it from the attention of a language model trained on reactions without annotations~\citep{schwaller2021}, GraphormerMapper applies a transformer to the molecular graphs~\citep{nugmanov2022graphormermapper}, and LocalMapper learns from reactions labelled by chemists~\citep{chen2024}. Astero and Rousu added the recognition of equivalent atoms to a graph-matching network~\citep{astero2024amnet,astero2025sammnet}. Each of these mappers returns one mapping for a reaction. Grandjean and co-workers recently compared RXNMapper and LocalMapper on three sets of reactions~\citep{grandjean2026}. They found that the measured accuracy depended on the set, and on whether two mappings that differ only by molecular symmetry were counted as the same. They also reported unreliable references in a validation set of United States patent reactions that is widely used to compare mappers. MAPPA uses no training data. The learned mappers are natural selectors for the alternatives that it returns.

\paragraph{Several mappings and molecular symmetry.} What should a mapper do when two mappings are equally good? Körner and Apostolakis gave an answer as early as 2008: keep every optimal mapping and discard those that differ only by symmetry~\citep{korner2008its}. On 1,542 annotated biochemical reactions, about 200 turned out to have more than one possible reaction center~\citep{apostolakis2008validation}. First, Gounaris and Floudas, and Latendresse and co-workers, reached the same answer with integer programming~\citep{first2012ilp,latendresse2012mwed}, and Mann and co-workers with constraint programming~\citep{mann2014cp}. Most recently, Ali and co-workers enumerated every mapping with the fewest bond changes and grouped them by symmetry~\citep{ali2025}. Their result is instructive. The correct mapping was among those returned for all 241 of their reactions, but it was the only one returned for just 151. Ties are therefore not a rare accident.

MAPPA owes its aim to this line of work, and it changes two things. These methods keep only the optimal mappings. We also keep those slightly above the optimum, because the mapping, especially between multi step reactions, are not always the one with the fewest bond changes (\Cref{fig:alternatives}). These methods also write out the mappings first and prune the equivalent ones afterwards. We never write them out: a family of mappings is stored as one combination of fragments, together with the rearrangements that symmetry allows.

\section{Methods}

\subsection{Overview: from shared fragments to alternative mappings}
\label{sec:architecture}
\label{sec:design-principle}

For each search direction, the \emph{base} is the endpoint on which fragments grow and the \emph{target} is where their placements are sought. Either endpoint can serve as the base; reversing the search exchanges these roles. MAPPA builds atom mappings by growing structural fragments shared between these endpoints (\Cref{fig:algorithm}). Starting from a seed atom, it extends a fragment while a compatible placement remains possible. When no admissible extension remains, it finds the completed fragment's alternative placements. Each placement defines a search branch: that choice is locked, and fragment growth continues on the remaining atoms under the accumulated assignments. Repeating this process produces alternative compositions of matched fragments. Different seeds, single-bond cuts and search directions expose additional compositions.

Two stages organize this procedure. The \emph{matching stage} discovers fragments and their conditional placements, then reduces repeated fragments and branch compositions into a shared mapping catalogue. The \emph{decoding stage} uses that catalogue to relate permitted atom rearrangements to bond events, preparing each unique fragment once and combining its response with the branch boundaries. Its output retains branches that can be queried within a chosen event-count window, rather than only one selected mapping. Matching and decoding are separate operations; a saved catalogue can be decoded without repeating discovery. Reference verification evaluates these outputs separately (\Cref{sec:verification}).

\subsection{Matching stage: grow, saturate and branch}
\label{sec:matching-rules}

Matching alternates between \emph{growth matching}, which tests for a relaxed subgraph isomorphism, and \emph{saturation matching}, which finds the completed fragment's distinct target atom-set placements. Postponing branching lets a small, common motif grow into a larger, more restrictive constraint (\Cref{fig:algorithm}a).

\subsubsection{Growth matching: keep extending while a match exists}
\label{sec:growth}

The growing fragment $g=(B_g,A_g)$ contains base atoms $B_g$ and accepted edges $A_g$; $\mathcal L$ records earlier locked assignments. A \emph{relaxed subgraph isomorphism} maps these atoms injectively into the target, preserving elements and locks, with each accepted edge placed on a target bond whose order differs from that of the base bond by at most the tolerance $\tau_{\mathrm{iso}}$. A \emph{witness} $m$ is one such map. Explicit hydrogens participate. Matching is non-induced: extra target edges are allowed, and unselected base bonds impose no additional constraint.

Growth starts from a seed atom that has an available target atom. A bond is an eligible connection when its weight in the target is at least the presence threshold $b$ (\Cref{eq:match}). MAPPA selects a boundary atom uniformly at random, then one of its frontier edges uniformly at random, and tentatively forms the extended fragment $g^+$. The decision is an existence test:
\begin{equation}
 \operatorname{Exists}(g^+\mid\mathcal L)
 =\mathbf{1}\!\left[\mathcal M(g^+\mid\mathcal L)\ne\varnothing\right],
 \label{eq:growth-exists}
\end{equation}
where $\mathcal M$ is the set of relaxed subgraph isomorphisms satisfying the locks. If a witness is found, MAPPA accepts the extension. If no placement exists, it defers that edge, keeps $g$ and tries the remaining frontier edges. \emph{Saturation occurs only when no eligible extension remains.} One failed edge does not end growth.

Let $\nu(g)$ count distinct target atom sets admitting a placement (\Cref{tab:growth-transitions}). Growth does not compute this count, because every positive outcome leads to the same decision: continue growing.

\begin{table}[H]
\centering\small
\caption{\textbf{Match-count transitions during proposed growth.} ``Many'' means more than one target atom-set placement, not multiple symmetry permutations on the same set. The $0\to0$ row is a rejected start; a zero-match fragment is never expanded.}
\label{tab:growth-transitions}
\begin{tabular}{@{}>{\centering\arraybackslash}p{.22\linewidth}>{\raggedright\arraybackslash}p{.72\linewidth}@{}}
\toprule
$\nu(g)\to\nu(g^+)$ & Decision\\
\midrule
$0\to0$ & No seed placement: skip this start and try another seed.\\
$1\to0$ & Keep $g$ and defer the edge. At saturation, one placement gives one continuation.\\
$\mathrm{many}\to0$ & Keep $g$ and defer the edge. At saturation, resolve its alternative placements.\\
$1\to1$ & Accept the extension and continue growing.\\
$1\to\mathrm{many}$ & Accept; postpone the placement choice until saturation.\\
$\mathrm{many}\to1$ & Accept; the larger fragment has narrowed the choices.\\
$\mathrm{many}\to\mathrm{many}$ & Accept; retain one witness and continue growing.\\
\bottomrule
\end{tabular}
\end{table}

For $1\to\mathrm{many}$, consider one eligible target carbon with two oxygen neighbors. C has one target atom set; C--O has two. Finding either placement accepts the extension without branching. The witness can change target sites during later growth, while earlier locks remain fixed. Extending the current witness or reusing a compatible saved fragment can avoid a new search; otherwise the existence solver finds one witness or establishes absence. A search that stops at its resource limit is not counted as a zero-match answer.

\subsubsection{Saturation matching: recover distinct atom-set placements}
\label{sec:completion}

At saturation, MAPPA finds distinct target atom sets $T_i=m(B_g)$ reached by relaxed subgraph isomorphisms. Each pair $(B_g,T_i)$ defines a completed fragment placement, with its internal mapping relation and the binary-graph automorphisms of its two sides retained compactly.

\begin{figure}[!htbp]
\centering\includegraphics{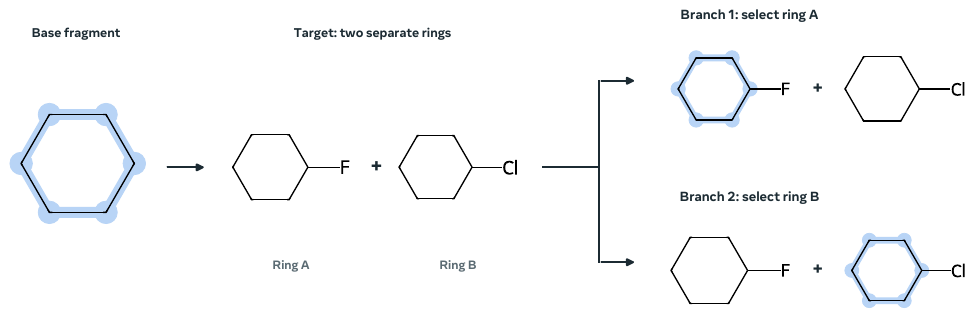}
\caption{\textbf{Two ring placements give two branches.} A fixed six-carbon fragment fits either target ring: the base atoms are the same, and the target atom sets differ. The different F and Cl attachments prevent a global automorphism from exchanging the rings. Each child selects one target carbon atom set, highlighted in blue; rotations and reflections within that ring remain inside the placement rather than creating additional branches. This is a carbon-fragment illustration with hydrogen constraints omitted, not a complete reaction.}
\label{fig:saturation}
\end{figure}

For example, A six-carbon ring fragment fitting two target rings with different external attachments gives two atom-set placements (\Cref{fig:saturation}). Ring rotations and reflections are fragment automorphisms: they change the correspondence within a placement without changing its atom sets. The search excludes target sets that are already represented, which avoids enumerating their bijections and removing the duplicates afterwards.

Such branching behavior allow effective consideration of different atom set to set relationships, allowing future searches to be more complete. In other words, fragments are only unique when they occupy different set of atoms between the base and the target.  

\subsubsection{Hierarchical branching: lock a placement and continue}
\label{sec:branching}

Each completed placement creates a child continuation, in which its assignments are locked and the next fragment is grown. In \Cref{fig:algorithm}a, choosing A$\to a$ leaves sites $b,c$ for B; choosing B$\to b$ leaves $c$ for C. These are conditional decisions: two fragments cannot independently occupy the same target atoms. Joining an earlier fragment retains its locked assignments and accepted constraints.

Each locked fragment expresses a shared structural constraint, extended as far as the current greedy trajectory permits. Another seed may discover a different constraint. 

However, we cannot gurantee that some seed are suboptimal, for example, think about having an independent H atom inside the base, and target has a lot of H groups. During early growth stage, when the seed land on the independent H atom, this will suddenly create significant amount of branches, that slow the search done due to high volume of parallel branches, while if the H atom is seeded in later search step, there aren't many atoms that left to be mapped. 

To address this potential inefficiencies, after each complete fragment growth step, MAPPA applies the branch cap from the leaves to the root: it discards the largest sibling subtrees as needed, recounts, and then assesses the parent. (\Cref{sec:hierarchy-details}). By pruning subtrees that hit the branch cap, we are effectively narrow the search into directions that leads to less sparsity, leaving high branching structure to be attended at later stage of search. However, this indeed is an empirical parameter to choose and might affect the final results. 

\begin{figure}[!htbp]
\centering\includegraphics{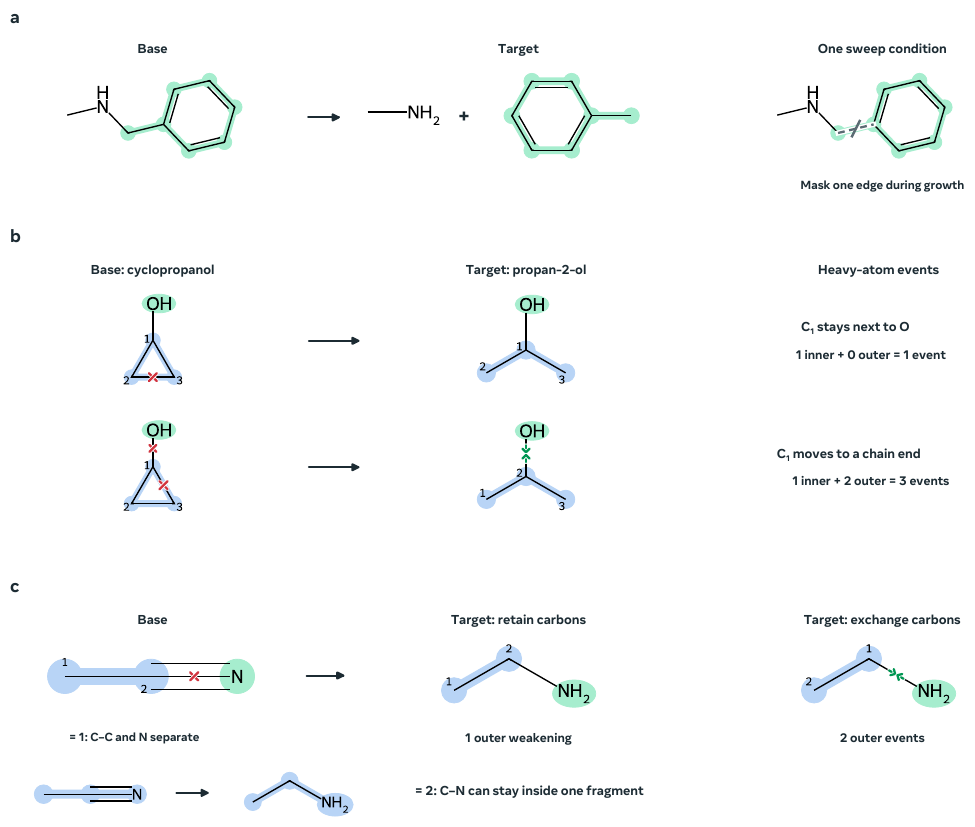}
\caption{\textbf{Molecular examples of exploration and boundary dependence.} (a) The benzyl region is preserved between N-methylbenzylamine and the displayed methylamine/toluene endpoint. Masking one of its edges changes the growth conditions; the grey mark is a search mask, not a claimed cleavage. (b) In cyclopropanol and propan-2-ol, the blue carbon fragment has one internal ring-opening event under either displayed assignment. Carbon 1 at the chain center preserves its oxygen attachment; at an end, it adds an outer cleavage and formation. The heavy-atom totals are one and three. (c) For acetonitrile and ethylamine, tolerance one can leave C--C and N separate: exchanging the carbons changes one outer weakening into two outer events. Tolerance two can include the changing C--N bond inside one fragment. Numbers track the same carbon atoms; red and green marks indicate losses and gains as in \Cref{fig:algorithm}. These are heavy-atom graph illustrations, with hydrogen constraints omitted and event threshold 0.5; panels b and c are checked by direct pair counting. They illustrate possible boundary dependence, not its measured frequency.}
\label{fig:design-examples}
\end{figure}

\begin{samepage}
\subsection{Matching stage: seeds, cuts and search direction}
\label{sec:cut-sweep}

\textbf{Vary the starting point.} One greedy trajectory can repeatedly favor a large preserved region and miss another useful decomposition. In \Cref{fig:design-examples}a, the highlighted benzyl region offers more starting atoms than the remaining methylamine region. Additional seeds can discover different constraints, but may also revisit the same large fragment.
\par\end{samepage}

\textbf{Vary the allowed connections.} A single-edge \emph{cut sweep} perturbs this commitment systematically. MAPPA searches the uncut graph and then repeats the search with each eligible base edge temporarily masked, including bonds to hydrogen. Each cut condition receives the requested number of seed orderings. Masking a connection inside the dominant region can expose a trajectory that ordinary seeding rarely reaches. The original endpoint weights remain unchanged for decoding: a mask alters discovery and does not assert a broken bond. Its endpoints may reconnect through another route, and one cut can redirect several later fragment choices.

\textbf{Exchange the endpoints.} Exchanging the base and target adds further alternatives. Seeds, cuts and reverse searches broaden discovery without making it exhaustive. The same exploration principle can be applied to SLAP, and \Cref{sec:method-coverage} reports its coverage with and without it.

\subsection{Matching stage: consolidate unique fragments and branches}
\label{sec:reduction}

Different seeds and cuts often rediscover the same fragment. The final matching subphase, \emph{reduction}, gives each base/target atom-set pair $(B_i,T_i)$ one shared identity $F_i$. For example, two searches that match $\{1,2,3\}$ to $\{a,b,c\}$ reference the same $F_i$, even if their growth order or starting witness differs. Each branch then references its fragment identities, allowing repeated compositions to share one catalogue entry.

Branches, which are composed by those unique fragments, are deduplicated by their unique fragments. This way, we retain unique mapping families after the search. 

\subsection{Decoding stage: prepare each fragment's symmetry--event response}
\label{sec:decoding}

Decoding turns the retained mapping alternatives into queryable bond-event information. It first prepares each unique fragment once, then reuses that response wherever the fragment occurs. The local calculation connects \emph{which atoms can move together} to \emph{which bond changes those moves produce}.

\subsubsection{Fragment automorphisms: independent and correlated moves}

A \emph{fragment automorphism} permutes atoms within one endpoint's fragment while preserving its binary connectivity and element labels, with anchors fixed. We compute these automorphisms separately for the base and target sides. Bonds satisfying the presence criterion share one structural label; original bond weights are retained for event counting. External attachments do not constrain the fragment automorphisms.

Nauty, a graph analtic package, supplies compact generators of these automorphism groups~\citep{mckay2014}: unit operations that describe the allowed rearrangements. Exchanging two identical arms moves each arm with its attached atoms; exchanging two hydrogens can act within an arm. The two arms must still occupy different positions. These joint operations retain that correlation without storing every full permutation. Both endpoint actions are represented; for an aligned equal-connectivity pair, one side suffices to describe their relative rearrangements.

\subsubsection{Internal events: attach bond changes to the allowed moves}

During matching, we use a large isomorphism tolerance for fragment discovery, but during decoding, we need to find if the previous permissive matching has internal bond events, such as bond weakening, and strengthening. In addition, possible automorphism permutations within the fragment might also produce distinctive bond events that needs to be investigated. Mathematically, let $W_B$ and $W_T$ denote the original base and target bond-weight matrices. For mapped atoms $u,v$, an assignment $m$ changes the bond weight by
\begin{equation}
 \Delta W_{uv}(m)=W_T(m(u),m(v))-W_B(u,v).
 \label{eq:local-event}
\end{equation}
The pair contributes one event when $|\Delta W_{uv}|$ exceeds its event threshold. We use empirical bond-order-change thresholds of 0.5 for organic bonds (the default for non-metal pairs) and 0.3 for bonds involving at least one metal atom. Note those metrics are feasible to be adjusted during practical scenarios. The sign distinguishes formation or strengthening from cleavage or weakening. These thresholds differ from the growth tolerance: a binary symmetry can preserve connectivity while changing bond weights.

In the acetate example (\Cref{fig:algorithm}c), O3 and O4 are interchangeable under an automorphism of the fragment's binary connectivity: each oxygen is adjacent to the same carbon. Their original C--O bond orders differ, so exchanging their assignments adds one weakening and one strengthening, giving two internal events. The saved response associates the rearrangement with those changes, so every branch using that fragment can reuse the calculation. Independent symmetry components are processed separately when no event term or requested endpoint couples them. Event-invariant choices share computation; correlated choices remain joint. The response supports later event queries without expanding every complete mapping.

\subsubsection{Different connectivity: improve the starting correspondence}

The accepted growth graph can omit bonds present in the final fragment. In the substituted triangle--chain example (\Cref{fig:design-examples}b), the displayed local shuffles keep one internal ring-opening event while changing the outer attachment. More generally, two valid relaxed subgraph isomorphisms on the same atom sets need not be connected by the final fragment automorphisms. Optimizing only one starting correspondence can therefore miss a lower-event alternative.

MAPPA first improves the saved witness under fragment automorphisms. Additional starting-correspondence search is allowed only when growth and final connectivity differ on at least one side and the witness has not attained an established event lower bound. Preparation and improvement share a fixed per-fragment budget. The output retains the selected starting correspondence and its symmetry response. An unsettled minimum is recorded as best found; selecting this starting orbit restricts decoded coverage. Solver details and bounds are given in \Cref{sec:decoding-details}.

\subsection{Decoding stage: assemble branch events and select the window}
\label{sec:window}

A branch combines three contributions: internal fragment events, ordinary boundary events and the correction for the swept bond. For a branch $\mathcal B$ and a permitted correspondence $m$, the reported count is
\begin{equation}
 E_{\mathcal B}(m)=B_{\mathcal B}+\sum_i e_i(m)+s_{\mathcal B}(m),
 \qquad 0\leq s_{\mathcal B}(m)\leq 2.
 \label{eq:composition}
\end{equation}
Here $e_i$ is the reused internal response of a disjoint final fragment piece, $B_{\mathcal B}$ is the ordinary-boundary contribution, and $s_{\mathcal B}$ accounts for the swept connection. The following rules specify how they are combined.

\subsubsection{Ordinary boundaries: a fixed contribution under the model}

MAPPA uses the independent ordinary-boundary approximation: these contributions remain fixed as internal fragment choices change. Bonds touching mapped atoms are counted, including mapped/unmapped boundaries; bonds wholly within unmatched material are omitted. No assignment is invented for an unmatched atom.

Greedy growth helps absorb connections that would otherwise couple fragments. In C--C$\equiv$N $\rightarrow$ C--C--N (\Cref{fig:design-examples}c), tolerance one can leave C--C separate from N. Swapping the carbons changes the outer contribution from one weakening to a cleavage plus a formation. Tolerance two allows C--N inside the fragment, where the internal response handles that dependence. Greediness helps, but does not guarantee independence: the triangle--chain example still has a variable outer attachment. Equation~\eqref{eq:composition} is therefore a model count, which can differ from the physical whole-reaction count.

Random growth, multiple seeds and the cut sweep can mitigate this limitation by discovering alternative fragment decompositions. In \Cref{fig:design-examples}b, growing through the O--C attachment first can place it inside a shared fragment instead of leaving it across a boundary. Another branch can then represent the same correspondence with its correct event count, even if the original branch assigns it differently. This exploration does not repair the original branch's approximation or guarantee a correct representation elsewhere. The benchmarks do not isolate the size of this mitigation.

\subsubsection{The swept bond: couple neighboring fragment responses}

A sweep masks one base bond. When its endpoints lie in different final fragments, those fragments are neighbors across the masked connection. Their internal event responses can be prepared separately, but this boundary contribution depends jointly on their choices: preserving the connection requires the two atoms to occupy compatible target positions. The swept bond therefore introduces a coupling between neighboring fragments that the decoder handles explicitly.

During branch assembly, MAPPA reuses the prepared responses to evaluate the two endpoint positions together, retaining their dependence on internal events and one-to-one occupancy. If both endpoints remain in one fragment, the same query uses that fragment's response. The associated contribution $s_{\mathcal B}$ ranges from zero to two events: a correspondence can preserve the connection, produce one bond change, or break one connection and form another. The sweep still masks only one bond; the event count describes the resulting correspondence across the two molecular endpoints.

\subsubsection{The event window: retain mappings near the minimum}

The event window focuses decoding on mappings with few bond changes. MAPPA first adds the local fragment minima and fixed boundary contribution for each branch. Let $L$ be the smallest of these totals before resolving the swept bond. When the fragment minima can be combined, the sweep adds at most two events, giving an initial branch count no greater than $L+2$. Only branches whose lower bounds are at most $L+2$ need to be examined for a lower count. Their swept-bond contributions are evaluated using the prepared fragment responses.

The lowest count found, $U$, sets the requested window $[U,U+n]$. MAPPA retains branches that could contribute mappings within this window and excludes any branch whose lower bound exceeds $U+n$. The studies here use $n=5$, and we call this choice the \emph{minimum-plus-five window}. Retained branches reference the shared fragment responses, allowing subsequent queries for mappings at a chosen event count or with specified atom correspondences.

\subsection{Execution: parallel stages and shared computation}
\label{sec:reuse-execution}

The stage boundaries also define parallel work. Reactions and directions are independent; seed/cut searches within a direction can run concurrently before their results are reduced together. Growth along one branch remains sequential because later fragments depend on earlier occupancy. For the reported batch workflow, the matching stage, including reduction, finishes before catalogue decoding begins. Different reaction/direction catalogues can then be decoded independently, with unique-fragment responses shared within each catalogue.

A persistent worker pool schedules these operations through the same interfaces used by direct local calls. Intermediate records can remain in memory or use aggregated storage. Compatible fragment proofs can be reused during growth, but each seed and cut keeps its own random decisions. Matching and decoding use a shared symmetry core, and bulk operations use native kernels. Scheduling and persistence do not determine which mappings are allowed. Algorithm details and pseudocode appear in \Cref{sec:algorithm-details}; execution and storage interfaces are documented in the repository.

\FloatBarrier
\section{Evaluation}

\subsection{Datasets and search settings}
\label{sec:current-validation}

We evaluated all 1,851 Golden reactions~\citep{lin2022} and the prepared collection of 47,974 EnzymeMap reactions~\citep{heid2023enzymemap}. Reference labels were removed before the search. Both studies used the prepared formal-bond-order matrices and explicit hydrogens. EnzymeMap supplies curated database mappings as references.

The common settings were discovery tolerance $\tau_{\mathrm{iso}}=2$, presence/cut threshold $b=0.2$, hierarchical branch cap 100, random seed 42, both directions and the full single-edge sweep. Golden was run with one and with ten seed orderings per cut, and EnzymeMap with one. Decoding used the original bond orders and the event thresholds given with \Cref{eq:local-event}, with a five-second allowance per unique fragment and the minimum-plus-five window. No reference mapping guided growth, pruning or starting-correspondence selection.

We compared one-seed MAPPA with two SLAP configurations on identical Golden inputs and the same CPU model. Standard SLAP used the uncut binary graph, and expanded SLAP combined binary and weighted modes with the full single-edge cut sweep. All three configurations ran in both directions. This paired study supplied the comparison of computational cost.

\subsection{Reference verification}
\label{sec:verification}

Coverage measures whether the annotated correspondence is present in MAPPA's decoded output within the minimum-plus-five event window. A recovered reference requires a concrete one-to-one mapping from a retained branch and its allowed symmetry choices. We replayed that mapping and checked its event count independently.

For both datasets, the check preserves annotated heavy-atom identities and unmatched-heavy-atom requirements; unannotated hydrogens must admit a feasible completion. Recovery in either search direction counts once. This measures reference inclusion among alternatives, not top-ranked prediction accuracy. Every reaction remains in the denominator. Comparator evaluations additionally allow chemically equivalent endpoint labels, preserving charge, isotope, hydrogen count and stereochemistry.

\subsection{Timing and reference-event analysis}
\label{sec:measurement}

Matching time includes both fragment discovery and reduction into the final catalogue; decoding is reported separately. CPU seconds sum successful worker computation over both directions and all cuts, including stage input loading and result construction. Publication and external verification are excluded. Source identities, hardware and completed-pass accounting are documented in the supplement.

To locate the reference relative to the detected minimum, let $m_{\mathrm{GT}}$ be a saved, verified mapping that agrees with the annotated atom pairs, the ground truth (GT). We measure
\begin{equation}
 n_{\mathrm{GT}}=E(m_{\mathrm{GT}})-U,
 \label{eq:reference-offset}
\end{equation}
where $U$ is the detected minimum already saved by decoding, taking the lower count across both directions. When both directions supply verified mappings, we use the one with the lower recorded count. Both terms use the same all-atom decoder model. Unannotated atoms, including hydrogens, retain their assignments in the verified witness. This comparison uses existing counts without further optimization.

For references absent from the saved output, we fixed their annotated pairs as anchors and completed the unspecified assignments with one seed per direction and no cut sweep. We compared the lowest attained completion count across both directions with the original $U$. The supplied pairs remained fixed, and the completed mapping was checked against the original bond matrices. These completion offsets were added to the same distribution, so that every reaction contributes once. Anchored completions do not contribute to independent mapping coverage.

We separately counted the distinct reaction cores observed among the saved mappings at the detected minimum. A core records which bonds decrease or increase in order, including bonds to hydrogen. Symmetry-equivalent atom labels count once, and different fragment partitions with the same bond-change pattern share one core. We combined both search directions in a common atom-index frame and evaluated changes directly from the original endpoint matrices. This measures the chemical differences between recorded alternatives without searching for additional symmetry arrangements.

\section{Results}

\subsection{MAPPA recovers more than 99\% of the references on both datasets}
\label{sec:reference-coverage}
\label{sec:method-coverage}

\begin{figure}[!htbp]
\centering\includegraphics{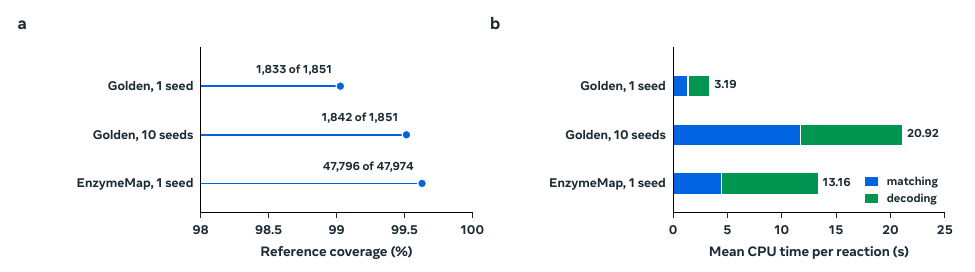}
\caption{\textbf{Reference coverage and computational cost.} (a) Reference inclusion in the decoded minimum-plus-five event window. Each label gives the references recovered and their total; the horizontal axis starts at 98\%. (b) Mean CPU seconds per reaction for matching (blue), including reduction, and decoding (green). The number beside each bar is their sum. Both directions and the full cut sweep contribute; publication and external verification are excluded.}
\label{fig:current-benchmarks}
\end{figure}

MAPPA recovered 1,833 of the 1,851 Golden references with one seed (99.03\%) and 1,842 with ten seeds (99.51\%). On EnzymeMap, one seed recovered 47,796 of 47,974 references (99.63\%; \Cref{tab:current-coverage,fig:current-benchmarks}). A reference counts as recovered when it occurs within the decoded minimum-plus-five window (\Cref{sec:verification}).

\begin{table}[H]
\centering\small
\caption{\textbf{Reference coverage on Golden and EnzymeMap.} A reference is recovered when a verified mapping occurs within the decoded minimum-plus-five event window in either direction. Percentages use all reactions.}
\label{tab:current-coverage}
    \begin{tabular}{lrrr}\toprule
Study & Reactions & Recovered & Not recovered\\\midrule
Golden, 1 seed & 1,851 & 1,833 (99.028\%) & 18\\
Golden, 10 seeds & 1,851 & 1,842 (99.514\%) & 9\\
EnzymeMap, 1 seed & 47,974 & 47,796 (99.629\%) & 178\\
\bottomrule\end{tabular}

\end{table}

Ten seeds recovered nine additional Golden references and lost none of those recovered by one seed. The wider exploration therefore improved coverage with the same decoding window.

MAPPA had the highest Golden coverage among the methods that we evaluated, 99.03\% with one seed (\Cref{tab:method-coverage}). A single prediction is treated as a candidate set of size one. Among the methods that return a single answer, LocalMapper had the highest coverage at 86.71\%, followed by Chython at 84.33\% and RXNMapper at 83.58\%~\citep{chen2024,chython,schwaller2021}. Default binary SLAP covered 85.95\%; combining binary and weighted modes with both search directions and the cut sweep raised its coverage to 96.92\% using our custom wrapper.

\begin{table}[!htbp]
\centering\small
\caption{\textbf{Golden reference coverage across AAM methods.} All methods are evaluated over 1,851 reactions. Coverage counts reference inclusion among returned alternatives; a single prediction is a singleton set. MAPPA uses the decoded minimum-plus-five output of the full one- and ten-seed studies. Reference checks follow \Cref{sec:verification}. Expanded SLAP includes both weight modes, both directions and the cut sweep.}
\label{tab:method-coverage}
\begin{tabular}{@{}lllr@{}}\toprule
Method & Search setting & Output & Coverage\\\midrule
\textbf{MAPPA} & \textbf{1 seed, sweep} & \textbf{Families} & \textbf{1,833 (99.03\%)}\\
\textbf{MAPPA} & \textbf{10 seeds, sweep} & \textbf{Families} & \textbf{1,842 (99.51\%)}\\
SLAP & Expanded sweep & Candidates & 1,794 (96.92\%)\\
SLAP & Binary, both directions & Candidates & 1,620 (87.52\%)\\
SLAP & Binary, default & Candidates & 1,591 (85.95\%)\\
LocalMapper & Default & Single mapping & 1,605 (86.71\%)\\
Chython & Default & Single mapping & 1,561 (84.33\%)\\
RXNMapper & Default & Single mapping & 1,547 (83.58\%)\\
RDT & Default & Single mapping & 1,030 (55.65\%)\\
Indigo & Default & Single mapping & 692 (37.39\%)\\
\bottomrule\end{tabular}

\end{table}

\subsection{Computational cost and seed scaling}
\label{sec:paired-slap}

One-seed MAPPA required an average of 3.19 CPU seconds per Golden reaction: 1.31 for matching and 1.88 for decoding. EnzymeMap required 13.16 CPU seconds, divided into 4.41 for matching and 8.75 for decoding (\Cref{tab:current-times}). The EnzymeMap median was 5.80 seconds and the 95th percentile was 41.86 seconds, so a minority of reactions accounts for much of the total computation.

\begin{table}[H]
\centering\small
\caption{\textbf{Mean CPU seconds per reaction.} Matching includes fragment discovery and reduction. Both directions and the full cut sweep are included. Decoding is measured separately; publication and external verification are excluded.}
\label{tab:current-times}
\begin{tabular}{lrrr}\toprule
Study & Matching & Decoding & Total\\\midrule
Golden, 1 seed & 1.31 & 1.88 & 3.19\\
Golden, 10 seeds & 11.64 & 9.28 & 20.92\\
EnzymeMap, 1 seed & 4.41 & 8.75 & 13.16\\
\bottomrule\end{tabular}

\end{table}

Increasing Golden from one to ten seeds raised the matching cost 8.87-fold and the decoding cost 4.95-fold. Matching scaled approximately linearly with seed count over this range. Decoding grew less because repeated fragments and branch compositions share their event responses. The additional exploration recovered nine more references at an average total cost of 20.92 CPU seconds per reaction.

\begin{figure}[!htbp]
\centering\includegraphics{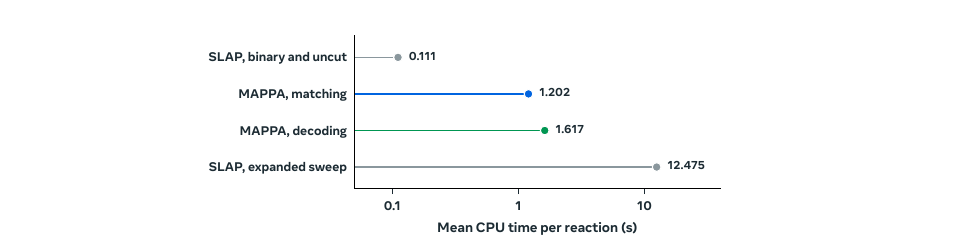}
\caption{\textbf{Computational cost under paired evaluation.} Mean CPU time on the 1,834 Golden reactions completed by all protocols, measured on the same CPU model. Standard SLAP is binary and uncut; expanded SLAP combines binary and weighted modes with the cut sweep. All protocols use both directions. MAPPA matching (blue) and decoding (green) are shown separately, and the two SLAP protocols are in grey; the time axis is logarithmic.}
\label{fig:paired-slap}
\end{figure}

The paired timing study compared computational cost on the same inputs and CPU model (\Cref{fig:paired-slap}). Standard binary SLAP was fastest. MAPPA required less computation than expanded SLAP, with its separate decoding stage included. Detailed measurement scopes are given in \Cref{sec:reproducibility}.

\subsection{About nine in ten references have the minimum event count}

\begin{figure}[!htbp]
\centering\includegraphics{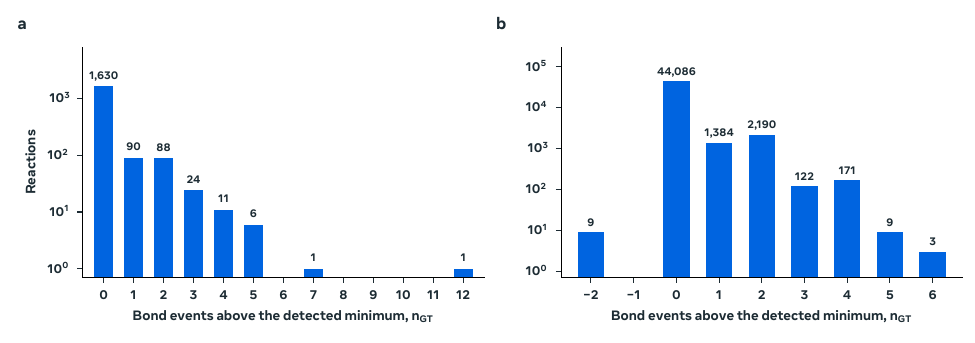}
\caption{\textbf{Reference-mapping event counts relative to the detected minimum.} Each reaction contributes $n_{\mathrm{GT}}=E(m_{\mathrm{GT}})-U$ (\Cref{eq:reference-offset}), using a verified reference-consistent full mapping and the original saved detected minimum. The lowest recorded verified count across both directions is used, with no additional optimization. Saved anchored completions supply the 18 Golden and 178 EnzymeMap references absent from the original output. Both terms use the same all-atom decoder model. Labels give reactions per bin; the vertical axis is logarithmic. The nine negative EnzymeMap offsets are anchored completions two events below the original detected minimum. Independent coverage is unchanged. (a) Golden, 1,851 reactions. (b) EnzymeMap, 47,974 reactions.}
\label{fig:reference-offsets}
\end{figure}

A low event count is a common criterion for choosing a mapping. We therefore asked how far the reference lies from the lowest count that MAPPA detected. The saved reference-consistent mappings had the same event count as the detected minimum for 1,630 Golden reactions (88.06\%) and 44,086 EnzymeMap reactions (91.90\%; \Cref{fig:reference-offsets}). Most other mappings lay within five events above the minimum. These distributions show that low event counts often agree with the reference, while retaining nearby alternatives provides additional reference-consistent choices.

Each offset describes a concrete reference-consistent mapping, including its saved hydrogen completion. It does not minimize over alternative completions of the reference.

Anchored completion succeeded for all 196 previously absent references, which are included in \Cref{fig:reference-offsets}. The completed count equalled the original detected minimum for nine Golden and 83 EnzymeMap cases, and nine EnzymeMap completions lay two events below it. The remaining completions lay above the detected minimum, reaching offsets of 12 in Golden and six in EnzymeMap. Every supplied pair was preserved, and each selected completion's model count agreed with direct counting from the original bond matrices. This figure shows that MAPPA very rarely miss the lowest bond event mapping, and also show case how necessary it is to preserve broad range of mapping instead of finding just mapping that reach the minimum events.

\subsection{One thirds of Golden reactions have more than one reaction core at the same bond events}
\label{sec:minimum-choices}

\begin{figure}[!htbp]
\centering\includegraphics{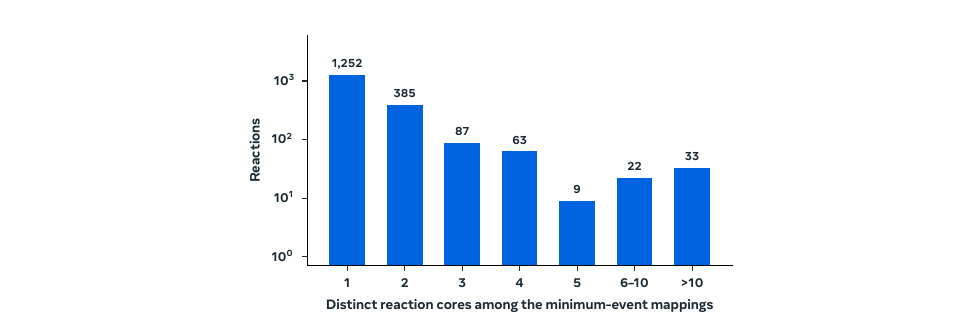}
\caption{\textbf{Distinct reaction cores among saved minimum-event mappings.} Counts cover all 1,851 one-seed Golden reactions and combine both search directions. Each core identifies bonds whose orders decrease or increase, including hydrogen events. Symmetry-equivalent atom labels count once, and different fragment partitions with the same bond-change pattern are merged. The vertical axis is logarithmic.}
\label{fig:minimum-choices}
\end{figure}

Several saved mappings can share the lowest event count, and different fragment compositions can describe the same bond changes. We therefore counted reaction cores, the patterns of bonds whose order decreases or increases (\Cref{sec:measurement}). Grouping the saved minimum-event mappings by their reaction cores gave a single core for 1,252 of the 1,851 Golden reactions (67.64\%; \Cref{fig:minimum-choices}). The median was one core, the mean 2.97 and the 95th percentile four. Most reactions therefore had few observed alternatives under this criterion, although a small tail reached 522 cores.

The remaining alternatives differ in their bond-change patterns, and not only in how the search partitioned the molecule. Equal event totals can thus retain distinct reaction cores for subsequent chemical selection. These are endpoint changes; they do not specify the sequence of elementary steps. This show cases that MAPPA can land potential equally good reaction patterns as alternatives.

\FloatBarrier
\section{Discussion}

MAPPA combines high reference coverage with explicit retention of mapping alternatives. Its fragment representation organizes these alternatives around shared structural constraints, allowing the same local relationship to contribute to many branches. Matching discovers those constraints, and decoding connects their permitted rearrangements to bond changes. This separation supports both the generation of candidate correspondences and the interpretation of how those candidates differ.

The seed study illustrates a practical trade-off between exploration and computation. One seed recovers most references, while additional seeds provide a modest coverage gain with approximately proportional matching cost. The cut sweep offers a complementary way to vary the constraints encountered during greedy growth. Together, these controls allow search effort to be adjusted to the application, with independent searches available for parallel execution.

The event distributions support using low-event mappings as a shortlist: the reference had the minimum count in 88\% of the Golden and 92\% of the EnzymeMap reactions (\Cref{fig:reference-offsets}). The minimum bond-event count alone is nevertheless not a reliable surrogate for the chemically correct mapping of an overall reaction. Overall reactions may comprise multiple elementary steps: the net bond changes between their endpoints do not describe the intervening mechanism, and the correct atom correspondence need not minimize those changes. Selecting only the smallest count can therefore discard a chemically meaningful higher-event mapping. MAPPA retains nearby alternatives so that reaction-specific chemical information can guide the final choice.

Retained alternatives could also assist reaction-path preparation. Atom correspondence determines how endpoint coordinates are paired during interpolation~\citep{de2019}; different assignments can therefore supply distinct initial paths for methods such as the nudged elastic band~\citep{henkelman2000}. Coupling these alternatives to chemical ranking and path refinement could help select correspondences using information beyond endpoint structure. Beyond pathway preparation, MAPPA could support reaction-template extraction from mapped bond changes. It could also serve as a structural verifier during reinforcement-learning training of large language models, providing feedback on proposed atom correspondences and bond changes.

\section*{Code and data}
Code and reproduction instructions are available at \href{https://github.com/yunhzou/MAPPA-AAM}{MAPPA source repository}, branch \texttt{main}. The repository guide \texttt{manuscript/REPRODUCIBILITY.md} links full benchmark configurations, software versions, per-case results, and figure-generation scripts. Golden and EnzymeMap are distributed by their original authors~\citep{lin2022,heid2023enzymemap}.
\par\noindent\begin{minipage}{\linewidth}
\section*{Acknowledgements}
We gratefully acknowledge the longstanding contributions of the Matter Lab’s current and past group members (matter.toronto.edu), and in particular those of the El Agente team (elagente.ca). C.C. acknowledges support from the DOE grant with the University of Minnesota, award \#A006801504, and the Basic Science Research Program through the National Research Foundation of Korea (NRF), funded by the Ministry of Education (RS-2025-02634334). Y.K. was supported by the CIFAR AI Safety Catalyst Award (Catalyst Fund Project \#CF26-AI-001). J.B.P.S. acknowledges funding of this project by the Natural Sciences and Engineering Research Council of Canada (NSERC) Alliance Grant \#ALLRP587593-23 (Quantamole). A.W. acknowledges support from the Canada Graduate Scholarships—Doctoral (CGS-D) program.
O.N.F. acknowledges support from the Spanish Ministry of Science, Innovation and Universities (MICIU) through a Salvador de Madariaga research mobility fellowship (grant PRX24/00448).

This research is part of the University of Toronto’s Acceleration Consortium, which receives funding from the CFREF-2022-00042 Canada First Research Excellence Fund. Computations were performed on the Niagara supercomputer at the SciNet HPC Consortium. SciNet is funded by the Canada Foundation for Innovation; the Government of Ontario; the Ontario Research Fund—Research Excellence; and the University of Toronto.

A.A.-G. thanks Anders G. Frøseth for his generous support and acknowledges the generous support of NSERC and the Canada 150 Research Chairs program. A.A.-G. and V.B. also acknowledge the generous support of Natural Resources Canada and the Canada 150 Research Chairs program and the University of Toronto’s Acceleration Consortium, which receives funding from the CFREF-2022-00042 Canada First Research Excellence Fund.

This work was supported by the Defense Advanced Research Projects Agency (DARPA) under Agreement No. HR0011262E022. This work was also supported by the AI2050 program of Schmidt Sciences.

Molecular graphics and analyses were performed with UCSF ChimeraX~\citep{goddard2018chimerax,pettersen2021chimerax}, developed by the Resource for Biocomputing, Visualization, and Informatics at the University of California, San Francisco, with support from National Institutes of Health R01-GM129325 and the Office of Cyber Infrastructure and Computational Biology, National Institute of Allergy and Infectious Diseases.

We thank Kjell Jorner and Robert Pollice for helpful discussions.

\par
\end{minipage}
\FloatBarrier
{\small\bibliography{references}\bibliographystyle{assets/plainnat}}
\clearpage
\appendix
\section{Technical algorithm details}
\label{sec:algorithm-details}
\setcounter{table}{0}
\renewcommand{\thetable}{A\arabic{table}}
\setcounter{figure}{0}
\renewcommand{\thefigure}{A\arabic{figure}}
\setcounter{algorithm}{0}
\renewcommand{\thealgorithm}{A\arabic{algorithm}}
\renewcommand{\theHalgorithm}{A.\arabic{algorithm}}

The complete stage sequence is summarized in \Cref{alg:mappa-overview}. The following sections specify the constraints, symmetry operations and event model used by this workflow.

\begin{algorithm}[!htbp]
\caption{Matching (including reduction), followed by decoding}
\label{alg:mappa-overview}
\small
\begin{algorithmic}[1]
\REQUIRE Endpoint graphs, discovery policies, seed/cut plan, branch cap and event window
\FORALL{requested search directions}
  \FORALL{cut conditions and seed orderings}
    \STATE Initialize the conditional search hierarchy
    \WHILE{a retained continuation has an eligible fragment seed}
      \STATE Grow using one-witness existence queries
      \STATE At saturation, construct distinct compressed placements
      \STATE Continue each placement with earlier assignments locked
      \STATE Prune and recount the hierarchy from leaves to root
    \ENDWHILE
    \STATE Retain terminal compositions and compact provenance
  \ENDFOR
  \STATE Share fragments by atom-set pair; reduce repeated branch compositions
\ENDFOR
\STATE Complete matching, including reduction, before batch decoding
\FORALL{reaction/direction catalogues}
  \STATE Inventory unique fragments and prepare their shared symmetry-event responses
  \STATE Compose branch contributions and the sweep-endpoint correction
  \STATE Retain queryable branches within the requested event window
\ENDFOR
\end{algorithmic}
\end{algorithm}

\subsection{Growth and completion constraints}
\label{sec:existence-details}

Let $V_B,V_T$ be the base and target atom sets, $Z_B,Z_T$ their elements, and $W_B,W_T$ their original bond-order matrices. Growth seeks an injective partial correspondence $m:D\subseteq V_B\rightarrow V_T$ that respects locked assignments. Every mapped atom preserves its element, and every accepted growth edge $(u,v)\in A$ satisfies
\begin{equation}
 Z_B(u)=Z_T(m(u)),\qquad W_T(m(u),m(v))\geq b,\qquad
 |W_B(u,v)-W_T(m(u),m(v))|\leq\tau_{\mathrm{iso}}.
 \label{eq:match}
\end{equation}
Only accepted growth edges belong to $A$. Nonedges and other bonds between the included atoms do not add matching constraints: this is non-induced relaxed subgraph isomorphism.

An existence query excludes occupied targets, removes destinations without compatible neighbors, and checks whether the remaining domains admit distinct targets. If these constraints do not decide the query, it searches the remaining assignments and stops at one witness. The same constraints establish absence when all choices are excluded. A witness may rearrange the growing fragment while earlier fragments remain locked. Compatible saved fragment witnesses can answer a query directly, provided they satisfy the current occupancy, anchors and accepted edges; reuse does not replace the new trajectory's random growth decisions.

\label{sec:completion-details}
At saturation, completion seeks distinct target atom sets, excluding each represented set from subsequent search. Internal assignments on one atom set remain compressed. A certified global target symmetry may additionally group placements when it preserves the entire remaining matching problem: the original weights, active connectivity, locks and discovery predicates. Otherwise the placements remain separate. Local fragment symmetry alone cannot establish that two global placements are interchangeable.

For each fragment side, nauty computes automorphism generators of its element-colored binary graph, with anchors fixed individually. Whole-ring or whole-arm moves remain correlated operations; terminal exchanges may be independent. Binary symmetry does not enforce bond-order tolerance. Discovery tolerance applies to the starting correspondence; subsequent local automorphisms on both sides are evaluated for their bond events using the original weights. Ordered global placement operations remain in their saved order.

\subsection{Hierarchical branching and fragment identities}
\label{sec:hierarchy-details}

Each completed placement creates a child continuation. After a completed step, MAPPA counts the surviving continuations from the leaves to the root. At a node over the limit, it discards the largest child subtree, recounts, and repeats until the retained count satisfies the cap; ties favor the earlier decision. The parent is then assessed using the reduced count. A separate completed-fragment limit rejects placement expansions exceeding its allowance. Internal symmetry permutations do not count as additional branches. Both limits restrict search coverage.

\label{sec:reduction-details}
The final matching subphase unifies fragments by their base/target atom-set pairs and assigns shared identifiers. One original fragment record is retained intact for each pair. Branches are deduplicated by their fragment composition and ordered placement context, while alternative cut and partition contexts are retained for decoding. This is structural deduplication, without symmetry optimization or event queries. Equal atom sets define a shared identity; they do not authorize arbitrary atom correspondences.

\subsection{Local event responses and branch composition}
\label{sec:decoding-details}

Each unique fragment is prepared once. Its response retains a selected starting correspondence, compact binary automorphism generators and the original-weight contributions to bond events. Operations are separated only when no variable event or required boundary endpoint couples them. An arm exchange therefore preserves the joint motion of that arm's atoms. Event-invariant operations share computation, while variable contributions retain their dependence on the relevant choices; the response does not store a Cartesian table of complete mappings.

Two relaxations provide lower bounds for local optimization. One compares bond-weight multisets within element-pair types without imposing atom consistency. The other matches incident bond assignments and then atoms injectively; because each internal event touches two atoms, half the incident cost, rounded up, is a lower bound. The stronger bound is used. These bounds need not be attainable.

MAPPA first improves the saved correspondence under its fragment automorphisms. It searches for another starting correspondence only when growth and final internal connectivity differ on either side and the current event count remains above an established bound. New starting correspondences must satisfy the saved discovery constraints. Subsequent binary symmetry choices are evaluated for events without reimposing discovery tolerance. Preparation and improvement share the per-fragment budget; attaining the bound ends optimization. Otherwise the response retains the best found witness and its bound. A minimum within the selected symmetry orbit is not a guarantee over every possible starting correspondence.

Branch assembly adds the disjoint fragment contributions and the fixed ordinary-boundary contribution. The single-edge sweep requires a joint correction for its endpoint choices, which can lie in neighboring fragments. That correction reuses the prepared local responses, preserves their dependence on internal events and contributes zero to two events. When internal costs are constant, only jointly reachable endpoint pairs need examination.

The decoder tracks each branch's lower bound and an attained count $U$. If compatible local minima attain the smallest pre-sweep composition $L$, the sweep correction gives $U\leq L+2$. Only compositions capable of improving $U$ need refinement; afterward, branches with lower bounds above $U+n$ are excluded. Unfinished local optimization retains its lower bound and an actual witness count rather than assuming the bound is attainable. Saved responses support subsequent queries within the window; a count interval does not imply that every integer inside it is attainable.

Ordinary boundaries use the independent-fragment approximation described in the main text. Consequently, the model event count can differ from a direct whole-reaction count. Reference-window membership uses the model count; physical bond changes are evaluated separately when constructing reaction cores.

\subsection{Molecular examples}
\label{sec:alternatives-provenance}

The introductory EnzymeMap example (case 13175, \Cref{fig:alternatives}) fixes two different heavy-atom correspondences: four heavy-atom events for the reference and two for the alternative. The endpoints contain nine heavy atoms and six hydrogens. Checking all $6!=720$ hydrogen assignments for each fixed heavy-atom correspondence gives six total events in both cases. For the reference, transfer of a carboxyl OH hydrogen contributes two additional events; for the alternative, transfer of both formaldehyde hydrogens contributes four. The example therefore shows distinct atom origins that remain tied after hydrogen completion.

The constructed matching diagrams illustrate graph relationships. In \Cref{fig:saturation}, the two target rings have different attachments and therefore define two distinct placements; each admits twelve internal rotations and reflections. Hydrogen constraints are omitted from this illustration. In \Cref{fig:algorithm}, exchanging the acetate oxygen images adds two internal events to two boundary events at the 0.5 threshold. These examples use formal bond orders. Their endpoint graphs and direct bond-change checks are supplied with the figure sources.

\section{Evaluation details}
\label{sec:reproducibility}

\subsection{Benchmark settings and reference checks}

The settings below apply to the current Golden and EnzymeMap studies. Golden uses one or ten seed orderings per cut; EnzymeMap uses one. Every search includes both directions, the uncut condition and all eligible single-edge cuts.

\begin{table}[H]
\centering\small
\caption{\textbf{Shared benchmark settings.} Events use the original bond orders; binary fragment graphs determine the allowed symmetry operations.}
\label{tab:benchmark-settings}
\begin{tabular}{ll}\toprule
Parameter & Value\\\midrule
Bond representation & Prepared formal bond orders, explicit hydrogens\\
Discovery tolerance $\tau_{\mathrm{iso}}$ & 2\\
Graph/cut presence threshold $b$ & 0.2\\
Hierarchical branch cap & 100\\
Random seed & 42\\
Bond-event threshold & 0.5 organic; 0.3 metal-related\\
Local preparation/improvement budget & 5 seconds per unique fragment\\
Decoded event window & $[U,U+5]$, with $U$ the best attained model count\\
\bottomrule\end{tabular}
\end{table}

An event requires a change strictly greater than its threshold. The empirical 0.3 threshold applies when at least one atom is a metal; 0.5 is the non-metal default. Reference checking preserves annotated heavy-atom pairs and unmatched-atom requirements. Unannotated hydrogen assignments must admit a complete injective witness, replayed through the saved response and independently counted inside the model window. The annotated reference does not guide the benchmark search.

Comparator outputs are evaluated over the same 1,851 Golden reactions, including invalid or failed outputs in the denominator. Single-answer methods contribute one candidate; SLAP contributes its returned candidates. Comparator checks permit chemically equivalent endpoint labels while preserving charge, isotope, hydrogen count and stereochemistry. Expanded SLAP combines binary and weighted modes, both directions and the single-edge sweep. Software versions, adapters, per-case outcomes and input identities are recorded in the repository's reproducibility guide.

\subsection{Timing scope}

Full-study CPU totals sum successful stage computation over both directions and every cut. Matching includes reduction; decoding is separate. These totals include stage loading and result construction, but exclude publication and external verification. Golden combines measured matching and postprocessing passes; EnzymeMap combines completed original tasks and completed recoveries. The totals measure accumulated computation, not uninterrupted elapsed time. Seed scaling compares the measured workloads rather than isolating a cache effect.

The paired timing comparison uses an Intel Xeon Gold 6442Y CPU and SLAPMapper revision \texttt{ea248fd}. All three protocols use identical prepared inputs, and their reported means use the same 1,834 completed reactions. These paired timings exclude loading, publication and external verification. Coverage retains all 1,851 reactions as its denominator. Hardware-controlled speed comparisons use this paired study, not ratios between the full Golden and EnzymeMap campaign means.

\subsection{Event offsets and reaction-core counting}

Each of the 49,825 reactions contributes one offset: the lowest saved verified reference-consistent model count across directions minus the original detected minimum. Unannotated hydrogen assignments are those of the selected witness. The 196 missing references use separately obtained anchored completions; those do not increase independent coverage. No branch-minimum or additional symmetry search is performed for the plot. The selected completion need not minimize over every possible hydrogen assignment.

Reaction-core counting uses saved witnesses at the detected minimum for all 1,851 one-seed Golden reactions. Both directions are converted to the same atom indexing. Direct comparison of the original matrices identifies signed bond changes, including hydrogen events; equivalent atom labels and different fragment partitions with the same pattern count once. This counts observed reaction cores in saved witnesses, without enumerating additional symmetry alternatives. A core describes net bond changes, not a complete sequence of elementary steps.

\begin{figure}[H]
\centering\includegraphics{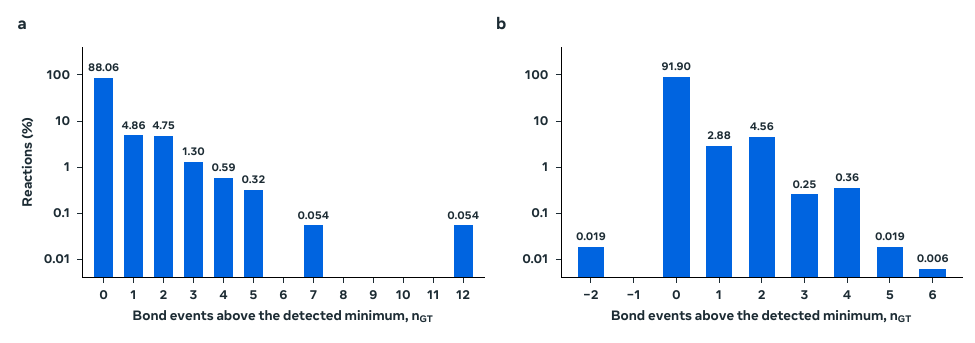}
\caption{\textbf{Reference event-offset distribution as percentages.} The same verified mapping counts and detected minima as in \Cref{fig:reference-offsets}, expressed as a percentage of all 1,851 Golden or 47,974 EnzymeMap reactions. Saved anchored completions are included. Each dataset sums to 100\% before rounding. Bar labels give percentages; the vertical axis is logarithmic. (a) Golden. (b) EnzymeMap.}
\label{fig:reference-percentages}
\end{figure}

\Cref{fig:reference-percentages} gives the percentage form of the event-offset distribution in \Cref{fig:reference-offsets}.

The repository's \texttt{manuscript/REPRODUCIBILITY.md} links the source versions, benchmark evidence, mapping certificates and figure-generation inputs supporting these analyses.

\end{document}